\documentclass[amsmath,amssymb,prl,twocolumn,superscriptaddress]{revtex4-2}
\usepackage[utf8]{inputenc}
\usepackage{graphicx}
\usepackage{dcolumn}
\usepackage{bm}
\usepackage{siunitx}
\usepackage[version=4]{mhchem}
\usepackage{natbib}
\usepackage{float}

\usepackage[shortlabels]{enumitem}

\usepackage{xcolor}

\newcommand{\beginsupplement}{
    \onecolumngrid
    \setcounter{table}{0}
    \renewcommand{\thetable}{S\arabic{table}}
    \setcounter{figure}{0}
    \renewcommand{\thefigure}{S\arabic{figure}}
    \setcounter{equation}{0}
    \renewcommand{\theequation}{S\arabic{equation}}
}

\begin{document}

\title{Evidence of Andreev blockade in a double quantum dot coupled to a superconductor}

\author{Po Zhang}
\affiliation{Department of Physics and Astronomy, University of Pittsburgh, Pittsburgh, PA, 15260, USA}

\author{Hao Wu}
\affiliation{Department of Physics and Astronomy, University of Pittsburgh, Pittsburgh, PA, 15260, USA}

\author{Jun Chen}
\affiliation{Department of Electrical and Computer Engineering, University of Pittsburgh, Pittsburgh, PA, 15260, USA}

\author{Sabbir A. Khan}
\affiliation{Microsoft Quantum Materials Lab Copenhagen, 2800 Lyngby, Denmark}
\affiliation{Center for Quantum Devices, Niels Bohr Institute, University of Copenhagen, 2100 Copenhagen, Denmark}

\author{Peter Krogstrup}
\affiliation{Microsoft Quantum Materials Lab Copenhagen, 2800 Lyngby, Denmark}
\affiliation{Center for Quantum Devices, Niels Bohr Institute, University of Copenhagen, 2100 Copenhagen, Denmark}

\author{David Pekker}
\affiliation{Department of Physics and Astronomy, University of Pittsburgh, Pittsburgh, PA, 15260, USA}

\author{Sergey M. Frolov}
\email{frolovsm@pitt.edu}
\affiliation{Department of Physics and Astronomy, University of Pittsburgh, Pittsburgh, PA, 15260, USA}

\begin{abstract}
We design and investigate an experimental system capable of entering an electron transport blockade regime in which a spin-triplet localized in the path of current is forbidden from entering a spin-singlet superconductor. To stabilize the triplet a double quantum dot is created electrostatically near a superconducting lead in an InAs nanowire. The superconducting lead is a molecular beam epitaxy grown Al shell. The shell is etched away over a wire segment to make room for the double dot and the normal metal gold lead.  The quantum dot closest to the normal lead exhibits Coulomb diamonds, the dot closest to the superconducting lead exhibits Andreev bound states and an induced gap. The experimental observations compare favorably to a theoretical model of Andreev blockade, named so because the triplet double dot configuration suppresses Andreev reflections. Observed leakage currents can be accounted for by finite temperature. We observe the predicted quadruple level degeneracy points of high current and a periodic conductance pattern controlled by the occupation of the normal dot. Even-odd transport asymmetry is lifted with increased temperature and magnetic field. This blockade phenomenon can be used to study spin structure of superconductors. It may also find utility in quantum computing devices that utilize Andreev or Majorana states.
\end{abstract}

\maketitle


Superconductor-semiconductor nanostructures are of interest in the context of quantum computing devices. Most prominently, they are a platform investigated as a host of Majorana zero modes that can be used to build topological qubits \cite{Kitaev2003, Nayak2008, Mourik2012, Deng2016, Alicea2011, Heck2012, Karzig2017}. While these qubits have not been achieved, other types of qubits namely Andreev, fluxonium and transmon have all been created out of super-semi structures \cite{delangeprl15, larsenPRL15,Hays2018, Luthi2018, Marta2020}. Spin qubits can also be hosted by quantum dots defined in semiconducting nanowires \cite{Nadj-Perge2010, Petersson2012, Berg2013}. Quantum dots exhibit iconic transport blockade phenomena: Coulomb blockade which is used in metrology to set quantum current standard \cite{Keller2008} and Pauli spin blockade which is used for readout and initialization of spin qubits \cite{Petta2005}. 


In quantum dots coupled to superconductors Andreev bound states form. They are hybrids of quantum dot energy levels and many-body Bogoliubov quasiparticles~\cite{Sauls2018, prada2020andreev}. Andreev double quantum dots have also been realized~\cite{Su2017, sherman2017normal, saldana2018supercurrent}. Here, we ask a question: can a blockade phenomenon unique to superconductors be demonstrated in Andreev quantum dots? Our goal is to observe Andreev blockade which is a suppression of Andreev reflection by a spin-triplet double dot configuration~\cite{Pekker2018}. 

\begin{figure}
\includegraphics[width=0.8\linewidth]{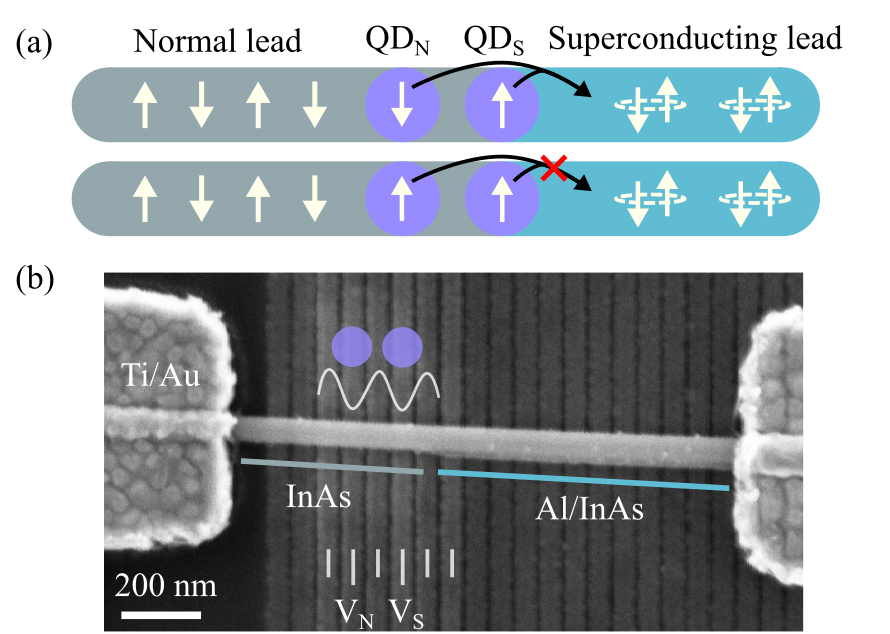}
\caption{\label{fig1}
(a) Schematic of Andreev blockade. The blockaded configuration is indicated with a red cross showing how a spin-triplet in double dot is prevented from forming a spin-singlet Cooper pair.
(b) Scanning electron microscope (SEM) image of a device similar to the one studied in the main text. Section marked 'Al/InAs' is an InAs nanowire covered by an Al shell. A section where the shell is etched is marked 'InAs'.  Vertical white lines mark gate electrodes used in creating the double dot.
}
\end{figure}


We fabricate a double quantum dot with one superconducting and one non-superconducting dot in an InAs semiconductor nanowire device (Fig. \ref{fig1}). The right dot exhibits an induced gap and Andreev bound states (see Figs.~\ref{fig2}(a) and (b)). The left dot is characterized by Coulomb diamonds and shows no dramatic superconducting features (see Figs.~\ref{fig2}(c) and (d)).  Low-bias (subgap) transport reveals patterns that theory predicted for the four-step Andreev charge transport cycle which arises when two electrons required to form a Cooper pair are transported through the double dot (Fig. \ref{fig3}). As evidence of Andreev blockade we find asymmetry between quadruple charge degeneracy points at even-to-odd and odd-to-even transitions in the normal dot (Figs. \ref{fig3} and \ref{fig4}). The observed asymmetry has the properties predicted by theory~\cite{Pekker2018}: the pattern is flipped at opposite voltage bias and it disappears at higher temperature and at higher magnetic field (Figs. \ref{fig5} and \ref{S2}). Experimentally, we find that current is not completely blocked in the regimes that we label as Andreev blockade. Our numerical model, based on Ref.~\cite{Pekker2018}, accounts for this by introducing finite temperature.

\section{Our Approach}

We experimentally realize conditions required for the observation of Andreev blockade following a theoretical proposal~\cite{Pekker2018}. We use a double quantum dot to trap a spin triplet state in an semiconductor nanowire (Fig.~\ref{fig1}(a)). The right side side of the double dot is connected to a superconductor, forming \ce{QD_S}, the left side is connected to a non-superconductor lead, such that \ce{QD_N} is a normal dot in the multi-electron regime.

Fig. \ref{fig1}(b) shows a scanning electron microscope image of the nanowire device. An InAs nanowire covered with $\sim$ 15 nm epitaxial Al is placed on top of 60 nm pitch gate electrodes. The double dot is defined electrostatically by voltages on gate electrodes indicated in the image. $V_N$ and $V_S$ are the gate voltages primarily used for tuning the dot states. Al on the left section of the wire is selectively etched to make the normal lead. The device is measured in a dilution fridge with a base temperature of about 40~mK.


\begin{figure}
\includegraphics{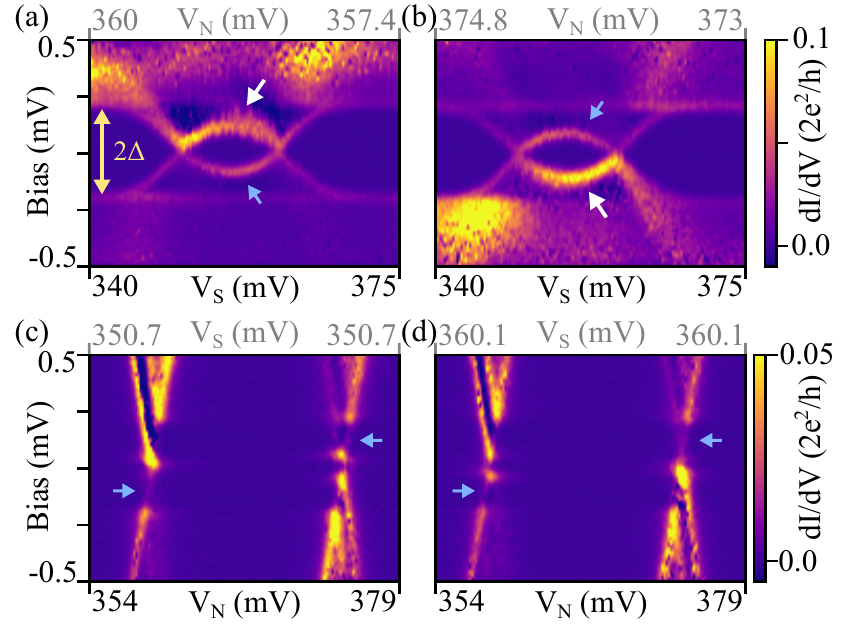}
\caption{\label{fig2}
Differential conductance spectra for (a-b) \ce{QD_S} and (c-d) \ce{QD_N}. Spectra are taken by fixing one dot at a degenerate state while tuning the other dot with the $(V_S, V_N)$ combination. The voltage combinations are indicated in Fig. \ref{fig3}(a). The large white arrows and small blue arrows indicate resonance peaks with different amplitudes. 
}
\end{figure}

\begin{figure}
\includegraphics{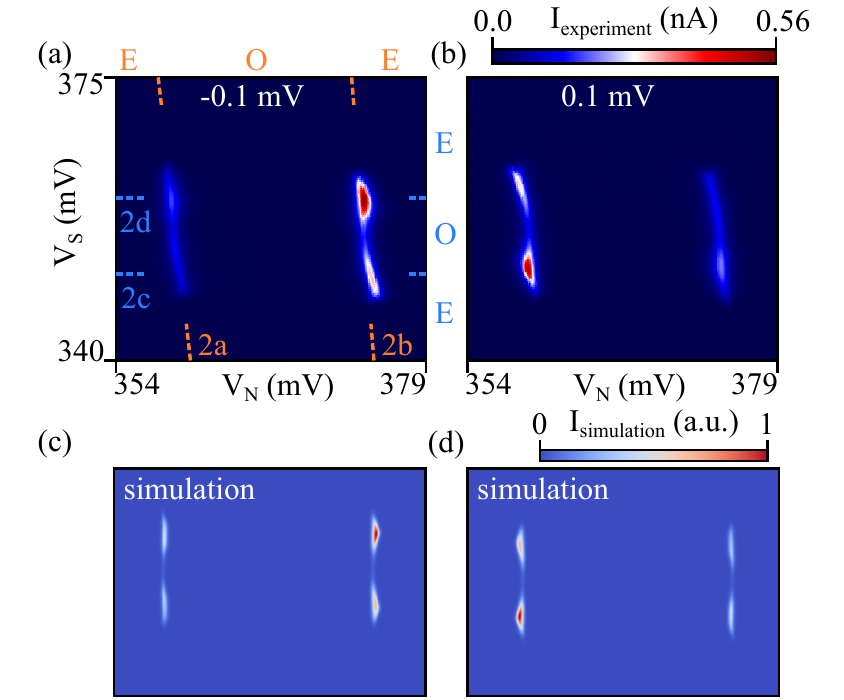}
\caption{\label{fig3}
(a-b) Experimental and (c-d) simulated results of a "unit" stability diagram with $2 \times 2$ quadruple degeneracy points. Parities in \ce{QD_N} and \ce{QD_S} are labeled on top and right axes in panel (a), respectively (E - even, O - odd). Dashed lines are traces along which spectra in Fig. \ref{fig2} are taken. The source-drain bias voltage is indicated in white. Parameters for simulation (in a.u. which match system energies in meV for convenience): source-drain bias $\mu_S-\mu_N =$ -0.1 in (c), 0.1 in (d), charging energy $U_N = 4$, $U_S = 0.7$, inter-dot charging energy $U_{NS} = 0.01$, induced gap $\Delta_S = 0.2$, temperature $T = 0.02$ .
}
\end{figure}

We demonstrate that dot $QD_S$ exhibits Andreev bound states, while dot $QD_N$ exhibits Coulomb diamonds, a staple of non-superconducting quantum dot transport (Fig.~\ref{fig2}). The double dot configuration is set up by tuning all of the electrostatic gates adjacent to the superconducting lead. Spectra are then taken by fixing one dot at a degeneracy point while tuning the other dot. Spectra of \ce{QD_S} show induced superconducting hard gap which is a stripe of suppressed current at voltage biases below $\Delta/e = 0.2$ mV, consistent with earlier works~\cite{Krogstrup2015, Chang2015, Gazibegovic2017, Khan2020}.  Inside the induced gap, loop-like resonances due to gate-tuned Andreev bound states are observed.  Spectra of \ce{QD_N} show Coulomb diamonds and no clear induced gap. To better understand the four panels of Fig.~\ref{fig2}, it is helpful to look at Fig.~\ref{fig3}(a) which shows a charge stability diagram and cuts in $V_N$-$V_S$ space that correspond to Fig.~\ref{fig2}.

There are also more subtle conditions that the system must meet for Andreev blockade to be observable. The induced superconducting gap should be hard in order to suppress single-particle transport below the gap. Any single-particle transport is an Andreev blockade lifting mechanism, thus the softer the induced gap the weaker are the Andreev blockade signatures. The barrier to the superconducting lead should be low in order to induce Andreev bound states. This is in contrast with Pauli blockade setups which typically require few electron regimes and hence high barriers to facilitate strong confinement. The inter-dot charging energy should be smaller than the induced gap because the Andreev transport regions shrink rapidly with the increasing of the inter-dot charging energy \cite{Pekker2018}. To match experimental results we set the inter-dot charging energy to 10~$\mu$eV, whic correponds to a weakly coupled double dot regime. 

\section{Predicted Signatures of Andreev blockade}

Following Ref.~\cite{Pekker2018}, we are looking for the following four experimental signatures (A-D) of Andreev transport and blockade in a N-\ce{QD_N}-\ce{QD_S}-S system.   We describe the charge sates in both dots by their parity, which can be either even or odd. The parity of states in \ce{QD_N} can be inferred by studying  how the degeneracy points shift in magnetic field, with odd regions expanding and even regions shrinking at higher fields (see Fig.~\ref{S7}). The parity of states in \ce{QD_S} can be inferred from Andreev spectra, with regions inside loop-like resonances being odd (see Fig. 2(a)-(b)). See supplementary materials and Ref.\cite{Pekker2018} for theoretical background underlying these signatures.

\begin{enumerate}[(A)]
\item
At subgap voltage biases current is confined to triangular regions of the charge stability diagram. The triangles do not appear in closely spaced pairs as in non-superconducting double dots where they form around triple points. Instead, triangles appear at quadruple Andreev degeneracy points, a consequence of the two-electron charge transfer cycle, form a parallelogram grid in $V_S$ vs $V_N$ space.
\item
An alternating pattern of blockade/no-blockade is observed when quadruple points are tuned by $V_N$. $V_S$ does not affect whether blockade is present or not. 
\item
The sign of source-drain bias voltage flips Andreev blockade. A quadruple point that is blockaded in positive bias is not blockaded in negative bias, and vice versa.

\item
Andreev blockade is not present when superconductivity is suppressed by magnetic field or temperature. Signatures (A)-(C) should no longer be manifest.

\end{enumerate}

Signatures (B) and (C) can be formulated together as follows. Andreev blockade is expected for (odd,odd)$\rightarrow$(even,even) charge parity transitions and for (odd,even)$\rightarrow$(even,odd) transitions, where the arrows indicate the direction of charge transfer, so that the conditions are valid for both signs of the applied bias.

An ideal blockade corresponds to total suppression of current below the gap. Blockade can be suppressed by the presence of sub-gap quasi-particles and thermally excited quasi-particles \cite{Pekker2018}. If Andreev blockade is only partially suppressed, a reduced current, known as leakage current, indicates blockade.

In principle there should be no fine-tuning required to observe Andreev blockade. All that is needed is one normal dot, one superconducting dot and a hard gap superconductor lead. Thus we are looking for a region of $V_S$ vs $V_N$ that includes many charge degeneracy points that exhibit blockade signatures. In practice, mesoscopic factors such as additional quantum dots in the nanowire segments covered by the leads can introduce their own current modulations. Thus some gate tuning may still be required to clearly observe Andreev blockade.

\section{Measured Andreev Blockade Signatures}

\begin{figure}
\includegraphics{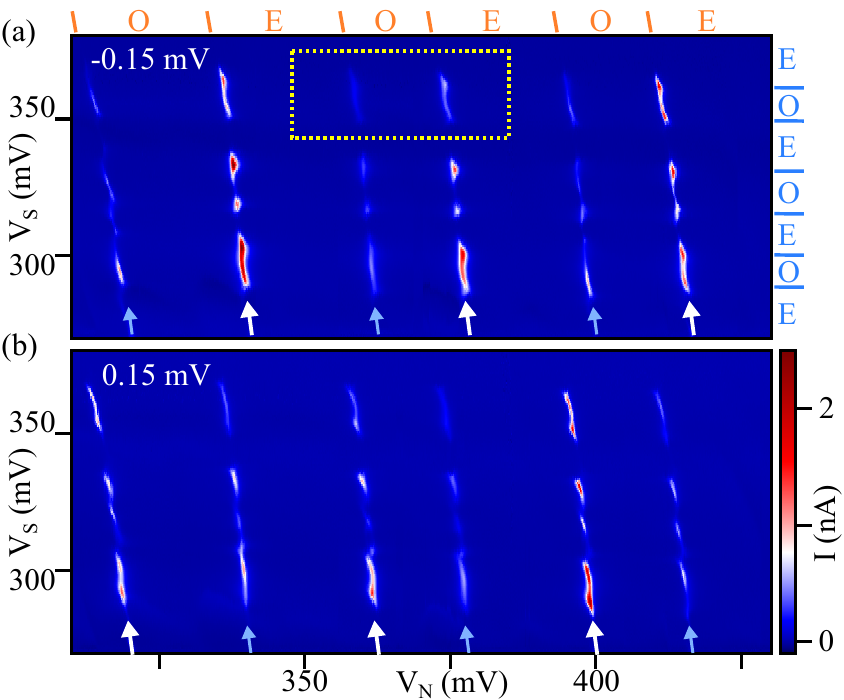}
\caption{\label{fig4}
Stability diagrams in a larger $(V_S, V_N)$ parameter space. The yellow dashed rectangle in panel (a) encloses the region studied in Figs. \ref{fig2}, \ref{fig3} and \ref{fig5}. Parities in \ce{QD_N} and \ce{QD_S} are labeled on top and right axes, respectively. Blue (white) arrows indicate columns of conductance triangles with low (high) current. The source-drain bias voltage is indicated in white.
}
\end{figure}

Signature (A) which is current confined to single, not double, triangles in the charge stability diagram is illustrated by Figs.~\ref{fig3}(a,b). Stability diagrams are taken at two opposite bias voltages. For both bias directions we observe elongated triangles, rounded due to relatively low bias voltages required to stay below the induced gap of aluminum. Numerical model results in Figs.~\ref{fig3}(c,d) closely reproduce the experiment. Larger gate range shown in Fig.~\ref{fig4} confirms the single-triangle character of the charge degeneracy points. 

Signature (B) in experiment presents itself as an alternating pattern of high current/low current when the occupation of \ce{QD_N} is changed. It is illustrated by Figs.~\ref{fig3} and \ref{fig4}. We see dim degeneracy points followed by bright ones. In Fig.~\ref{fig4} the dim columns are marked by dim arrows, while the bright columns are marked by bright arrows. The region of $(V_S, V_N)$ parameter space depicted contains $6 \times 6$ degenerate points. All degeneracy points are detectable, even those that are supposed to be blockaded. In the context of Andreev blockade this means that the blockade is partially lifted. In the simulation of Figs. \ref{fig3}(c,d) we assume finite temperature to reproduce this behavior. Finite temperature enables single particle tunneling into a hard-gap superconducting lead via thermally excited quasi-particles, and provides a way around the blocked Andreev processes.

Signature (C) is the reversal of the high/low current pattern in opposite bias. In Fig.~\ref{fig3}(a), at $-0.1$ mV, current is smaller for the degeneracy points on the left. In Fig.~\ref{fig3}(b), at $+0.1$ mV, the current is smaller for the degeneracy points on the right. Simulated Andreev blockade regime shows good agreement with this observation (see Figs.~\ref{fig3}(a),(b)). The same behavior largely holds in Figs.~\ref{fig4}(a) and (b) over an expanded range of gate voltages covering 6$\times$6 degeneracy points. 

Bias asymmetry can also be seen in Fig.~\ref{fig2}. In the upper panels, Andreev loops have non-symmetric amplitudes between positive and negative bias voltages marked by large and small arrows: either the upper or the lower half of the loop is brighter than the other half. In Coulomb diamonds shown in the lower panels, the pattern of intensity is anti-symmetric with respect to the center of the figure. For example, in panel \ref{fig2}(c), the left region is bright at positive bias, while the right region is bright at negative bias - when looking at biases below 0.2 mV, the induced gap (also marked by arrows). These patterns are consistent with Andreev blockade: when the occupation of \ce{QD_S} is changed AB is not affected, and when the occupation of \ce{QD_N} is changed AB appears at the opposite bias. The full picture is more complicated as certain bias asymmetries are also observed at currents above the induced gap. However, Figs. \ref{S3} and \ref{S4} illustrate that in general bias asymmetry at high bias does not follow the same pattern as subgap low-bias asymmetry, suggesting that they are of different origin. 

\begin{figure}
\includegraphics{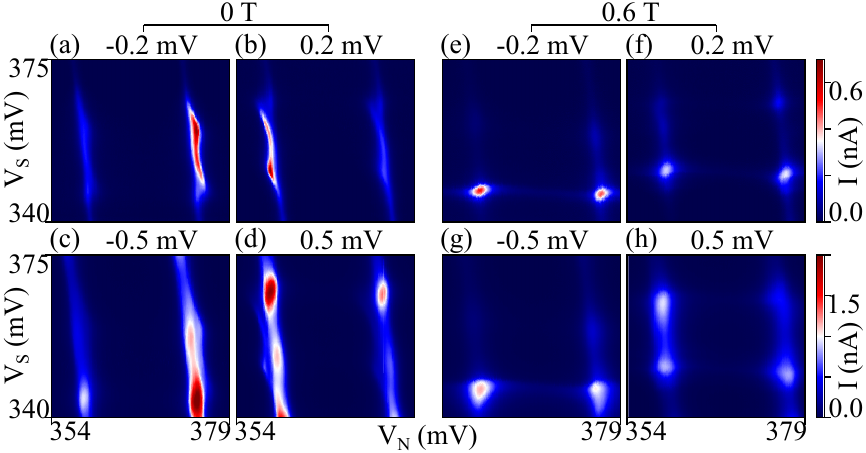}
\caption{\label{fig5}
Stability diagrams at different bias voltages and magnetic fields (a-d) $B$ = 0 and (e-h) $B$ = 0.6 T. The bias voltage is indicated in each panel. The magnetic field is in the sample plane at a 24 degree angle with the nanowire.
}
\end{figure}

Finally, signature (D) which is the disappearance of other signatures when superconductivity is suppressed is presented in Fig. \ref{fig5} and supplementary figures \ref{S2}, \ref{S6}, \ref{S9}, \ref{S10}. Figs. \ref{fig5}(a-d) reproduce the same regime as in Fig.~\ref{fig3} for different bias voltages. When a magnetic field of 0.6 T is applied, in Figs. \ref{fig5}(e-h), we observe that the alternating patterns of bright/dim degeneracy points are no longer present, and neither is the bias voltage asymmetry. The same holds true at elevated temperature, as illustrated in Fig. \ref{S2}.

Single elongated degeneracy points are replaced by less elongated points at higher fields where superconductivity of the aluminum shell is suppressed (Figs. \ref{fig5}(e-h)) - in this regime the charge degeneracy points appear more similar to those of a normal double dot with pairs of triangular triple points, but with significant rounding and blurring and a weak interdot capacitive coupling. In future experiments using a larger gap superconductor such as Sn or Pb \cite{Pendharkar2019, Kanne2020, Khan2020} as a shell can provide a larger bias voltage range for the observation of Andreev blockade and make this observation more clear by reducing the relative role of feature broadening.

\section{Alternative explanations}

We have considered the possibility that signatures (A-D) were only identified due to fine-tuning in a deliberate search for predicted patterns. In this scenario, signatures such as alternating bright/dim degeneracy points and bias asymmetries are not due to Andreev blockade, but rather they arise accidentally due to additional states co-existing with the \ce{QD_N}-\ce{QD_S} system in the same nanowire - for example spurious Andreev and normal quantum dots in the nanowire lead segments. Those other states are fine-tuned to modulate transport in the double dot in just the right way to be consistent with Andreev blockade.

We cannot fully exclude the possibility of the presence of extra mesoscopic states beyond the two quantum dots. We see, e.g. Fig.~\ref{fig4} lower left part, that while the stability points form a dominating double-dot pattern, their intensities vary across a large $V_S$-$V_N$ range suggesting non-monotonic coupling to states outside the dots or non-monotonic inter-dot barriers. This is typical for a wide variety of quantum dots, including those made in the Intel cleanroom \cite{zwerver2021qubits}.

Additional data in supplementary materials and full experimental data on Zenodo \cite{zenodo_ABexp} also demonstrate that within the same device other regimes do not show patterns of Andreev blockade, when all gates used to define a double dot are set differently. 

Our argument to not favor the above explanation is that a pattern consistent with Andreev blockade signatures (A-C) are observed over a regime covering 6$\times$6 degeneracy points, in several sufficiently different double dot configurations, and, in a more limited range in other devices. Furthermore, the fact that bias asymmetry and alternating current patterns disappear when superconductivity is suppressed (signature D) convinces us that these phenomena have to do with subgap superconducting transport which is the regime of Andreev reflection.

\section{Follow-on Work}

Recently larger hard gaps have been induced in semiconductor nanowires in experiments with tin (gap of 0.6 mV) and lead (gap of 1.2 mV) shells \cite{Pendharkar2019, Kanne2020}. It would be interesting to repeat Andreev blockade experiments using these superconductors. First, larger ratio of gap to measurement temperature may result in stronger blockade. Second, the ability to work at higher bias and larger charging energies would make the observation of various blockade features such as bias triangles more conclusive, and reduce the role that rounding plays at low biases. Finally, blockade can be studied to higher magnetic field allowing for a detailed investigation of spin structure in the parent superconductor.

Several improvements can be done in immediate follow-on work related to materials processing and device fabrication. This would impact not only Andreev blockade experiments but many works aimed at searching for Majorana modes and building superconductor-semiconductor qubits. For instance, the wet etch degrades the quality of nanowires by introducing defects. The supplementary information in Ref. \cite{Carrad2020,Khan2020} shows that even using the standard etchant for InAs-Al wires may result in InAs damage or Al islands. The use of in-situ shadowing or dry etching are promising avenues to explore.

\section{Future relevance}
At the most basic level, Andreev blockade offers a means of studying spin-resolved transport in hybrid devices at zero magnetic field. We foresee application of Andreev blockade in experiments that probe spin pairing in superconductors. Much like Pauli blockade was used to investigate spin mixing mechanisms in semiconductors due to hyperfine, spin-orbit or electron-phonon coupling, Andreev blockade can be potentially used to detect triplet pairing or admixtures thereof, spin-flip scattering, spin polarization or textures such as Larkin-Ovchnnikov-Fulde-Ferrel state in the superconductor. A two-arm Andreev blockade device with two double dots in parallel can in principle be used as a spin-sensitive probe for crossed Andreev reflection. Quantum dots with superconducting leads are building blocks of Andreev qubits, Kitaev emulators and of topological qubits \cite{Hays2018, Sau2012, Karzig2017}. These devices may manifest Andreev blockade or utilize it to detect the state of a qubit or an emulator by providing a spin-dependent transport or transition rate element.

\section{Comparison to other works}

Several versions of a triplet blockade in quantum dots closely related to Andreev blockade have been considered theoretically~\cite{Choi2000, Recher2001, Padurariu2012, Droste2012, Tanaka2010}, with several works focusing on a parallel combination of quantum dots, which is relevant for crossed Andreev reflection~\cite{Eldridge2010, Leijnse2013, Scheruebl2019}. Other types of blockade related to Andreev reflection such as chiral blockade have been proposed~\cite{Bovenzi2017}. 

A recent experiment in a similar double dot setup with two rather than one superconducting lead has studied a triplet blockade that develops at large magnetic field, where spin triplet is the unique ground state of the double dot~\cite{Bouman2020}. In contrast, Andreev blockade demonstrated in this work occurs at zero magnetic field, due to the stochastic filling of a quantum dot by random spins.

\section{Funding Sources}
S.F. and D.P. are supported by NSF PIRE-1743717. S.F. is supported by NSF DMR-1906325, ONR and ARO. P.K. is supported by European Union Horizon 2020 research and innovation program under the Marie Skłodowska-Curie Grant No. 722176 (INDEED), Microsoft Quantum and the European Research Council (ERC) under Grant No. 716655 (HEMs-DAM).

\section{References}

\bibliographystyle{apsrev4-1}
\bibliography{references.bib}
\clearpage

\beginsupplement

\begin{center}

\textbf{\large Supplementary Materials: Evidence of Andreev blockade in a double quantum dot coupled to a superconductor}

\vspace{0.45 cm}
\end{center}

\section{Theoretical background: Andreev blockade}

Theory of Andreev blockade and the details of our numerical model are contained in a separate paper~\cite{Pekker2018}. Here we summarize key concepts that are relevant for our interpretation of the experimental data.

Andreev blockade is a consequence of a dynamical formation of a stable sub-gap spin-triplet state. The series double dot setup allows for this to happen: a triplet state with one spin on each dot can have a low chemical potential due to small exchange interaction between spins. 

Andreev blockade is the suppression of Andreev reflection due to spin parity mismatch between the double quantum dot and the superconductor. Andreev blockade is most apparent when the superconductor induces a hard gap in the nanowire. The hard gap ensures the suppression of single-particle tunneling and enforces Andreev reflection as dominant means of transport. 

Andreev reflection transfers a charge of 2e into the superconductor, where 'e' is the electron charge. In order to move the two electrons between the normal and the superconducting lead, the double dot transitions through four charge configurations per Andreev cycle. Therefore transport is only allowed at quadruple charge degeneracy points where four charge configurations have similar chemical potentials all within a source-drain bias window that does not exceed the superconducting gap. As a point of comparison, in non-superconducting double dots transport involves moving just one electron between the leads and takes place at triple, rather than quadruple, degeneracy points, which results in the formations of the well-known honeycomb charge stability diagram~\cite{Hanson2007}.

Andreev blockade is controlled by the occupation of the normal dot \ce{QD_N}, but it is not sensitive to the occupation of the superconducting dot \ce{QD_S}. Quantum dot \ce{QD_S} is strongly coupled to the superconductor, and this coupling hybridizes all states of the same parity. For example, states with 0 and 2 charges are hybridized, and so are states with 1 and 3 charges. Coupling to the superconductor imposes an approximate particle-hole symmetry which mandates that transport at odd-to-even and even-to-odd degeneracy points is the same. This translates into the insensitivity of Andreev blockade to the occupation of \ce{QD_S}. 

A complimentary way to see why transport is insensitive to the charging state of \ce{QD_S} is as follows. Over the course of a transport cycle two electrons must be added to \ce{QD_S}, driving the dot from an even parity state, to an odd parity state, back to an even parity state. If the normal dot \ce{QD_N} has two charges on it, then they must be of opposite spins and one of them can always escape into \ce{QD_S}. Specifically, if \ce{QD_S} is in the even parity state, either of the spins on \ce{QD_N} can move to \ce{QD_S} thus putting \ce{QD_S} into an odd parity state. If \ce{QD_S} is in an odd parity state, one of the electrons on \ce{QD_N} has the opposite spin and can move to \ce{QD_S} resulting in a transition to an even parity state. On the other hand, if \ce{QD_N} only has one spin, then that spin cannot escape to \ce{QD_S} if \ce{QD_S} is an odd parity state of the same spin. Hence, the charging state of \ce{QD_N} determines if an Andreev blockade is established. 

\section{Pauli blockade vs. Andreev blockade}

Both Pauli blockade and Andreev blockade occur in double quantum dots with two charges in spin triplet state. However, they have different origins. Pauli blockade is due to Pauli principle which prevents two electrons of the same spin from occupying the same orbital. Andreev blockade is due to inability of forming Cooper pairs out of spin-triplet pairs. The characteristic energy scale of Andreev blockade is the induced superconducting gap. The characteristic energy scale of Pauli blockade is the singlet-triplet energy level spacing in the (0,2) charge configuration.

Andreev and Pauli blockades appear in different yet overlapping parameter spaces and can be observed in the same device. In soft gap Andreev double quantum dots, Pauli blockade has been previously observed in Ref. \cite{Su2017}. Though we did not observe Pauli blockade in the device studied here. This is consistent with what is known of Pauli blockade - it is a relatively rare phenomenon which is observed more frequently in few-electron double dots, due to larger singlet-triplet energies. Whereas, in multi-electron dots, such as those studied here, Pauli blockade appears in one out of every 10 or 100 degeneracy points \cite{Zarassi2017}. In contrast, Andreev blockade is not expected to be as sensitive to quantum dot energy scales and is, in principle, guaranteed by the superconducting gap. Even if singlet-triplet energies were significant for all orbitals, Pauli blockade would be expected in 1 out of 4 degeneracy points, while Andreev blockade is expected in 2 out of 4.

\section{Further reading}

Background information on non superconducting single and double quantum dots can be found in Ref.~\cite{Hanson2007}. An introduction to Andreev bound states can be found in Refs.~\cite{Sauls2018, prada2020andreev}. A related subject of Majorana zero modes in 1D wires is also discussed in Ref.~\cite{prada2020andreev}. Experiments on epitaxial growth of Al on nanowires and the hard gap can be found in Ref.~\cite{Krogstrup2015, Chang2015}.

\section{Methods}

The growth of InAs nanowires with Al shells are performed using molecular beam epitaxy (MBE). First, InAs nanowires are grown from predefined Au catalysts via vapor-liquid-solid mechanism. After the nanowire growth, the growth chamber is cooled down and \textit{in situ} Al growth is carried out, ensuring the high interface quality of the hybrid. Further discussion about the growth can be found in Ref. \cite{Khan2020}.

Electrostatic 60 nm pitch gates are patterned by 100 kV e-beam lithography (EBL) with PMMA 950 A1 as the resist. The development for gate patterns is performed in 1:3 MIBK/IPA for 1 minute at a low temperature of -15 $^\circ$C to enhance the resolution \cite{Cord2007}. Then a bilayer of 1.5/6 nm Ti/PdAu is evaporated by electron-beam evaporation. The gates are covered by 10 nm of \ce{HfO2} as the dielectric layer, patterned by EBL and grown by atomic layer deposition (100 cycles).

Nanowires are transferred onto the gate chip using a micro-manipulator under an optical microscope. Windows for Al-etching are defined by 10 kV EBL. The PMMA resist is dried in a vacuum chamber at room temperature to avoid heating the Al layer \cite{Gazibegovic2017}. \ce{AlO_x}/Al on the InAs wire is selectively etched using MF CD-26 developer/DI water solution for 2 minutes, at a volume ratio of 1:20 at room temperature. The etching of the shell over a significant length of the nanowire (hundreds of nanometers) is a delicate process which requires optimization. If not optimized, residual grains of aluminum are left on the nanowire and they are capable of inducing superconductivity in the regions where this is not desired, e.g. over the dot $QD_N$.  

Leads are made with EBL, Ar cleaning, followed by e-beam evaporation of 10/130 nm of Ti/Au.

Measurements are performed in a dilution refrigerator with a 40 mK base temperature.

\section{Author contributions}
S.A.K. and P.K. grew the nanowires. J.C., H.W., and P.Z. fabricated the devices. D.P., P.Z. and S.M.F. developed the theory and performed numerical simulations. P.Z. and H.W. performed measurements. P.Z. and S.M.F. wrote the manuscript with inputs from all authors.

\section{Volume and duration of study}

The first InAs-Al device for the transport component of the study was measured in March 2019, when we performed characterizations of newly arrived wires and tested fabrication recipes. A total of 8 double dot devices were studied. The device discussed in this manuscript was made in March 2020 and measured from June to September 2020, producing 6855 datasets. The interruption between the fabrication and the measurement is due to shutdown caused by COVID-19.

\section{Data availability}
Curated library of data extending beyond what is presented in the paper, as well as simulation and data processing code are available at \cite{zenodo_ABexp}.

\section{Supplementary data for the device A presented in the main text}

\begin{figure}[H]\centering
\includegraphics{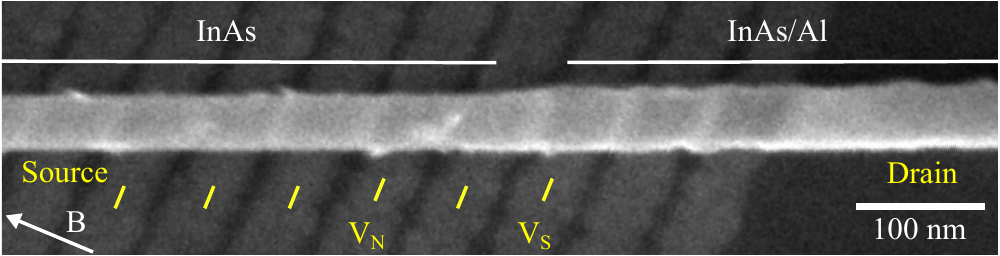}
\caption{\label{S1}
SEM image of device A that corresponds to data in the main text. The image is taken after Al etch but before the evaporation of Ti/Au leads. The double quantum dot is created electrostatically using 6 gate electrodes (yellow lines) and the underlying doped Si substrate (as a global gate).
}
\end{figure}

\begin{figure}[H]\centering
\includegraphics[scale=1.2]{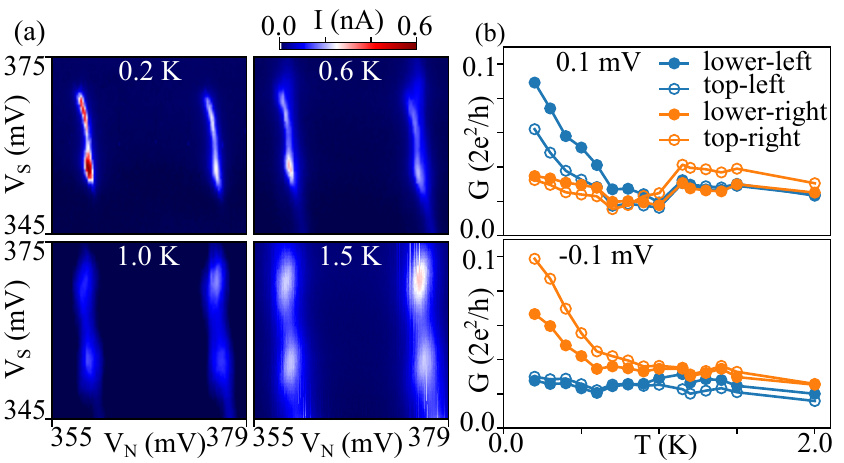}
\caption{\label{S2}
Temperature dependence of the transport region in Fig. 
\ref{fig3}.  (a) Stability diagrams at different temperatures.  The temperature is labeled in each sub-panel. The source-drain bias voltage is 0.1 mV. (b) Local maximum conductance extracted near four degeneracy points versus the temperature. The asymmetry between the left and right maxima disappears above Al's bulk critical temperature (1.2 K). B=0.
}
\end{figure}

\begin{figure}[H]\centering
\includegraphics{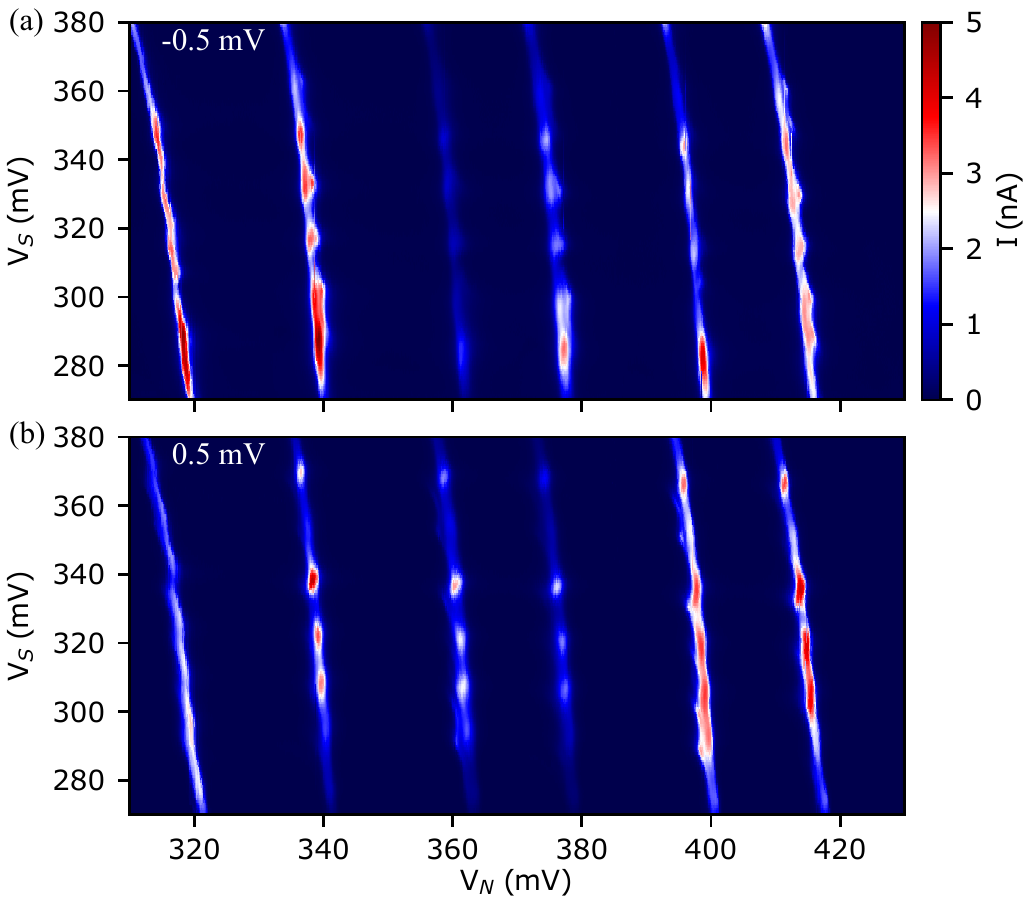}
\caption{\label{S3}
Additional data to Fig. \ref{fig4}, at bias voltages above the gap. The source-drain bias voltage is indicated in white. B=0, base temperature. The even/odd pattern of high/low current is not present at high bias.
}
\end{figure}

\begin{figure}[H]\centering
\includegraphics{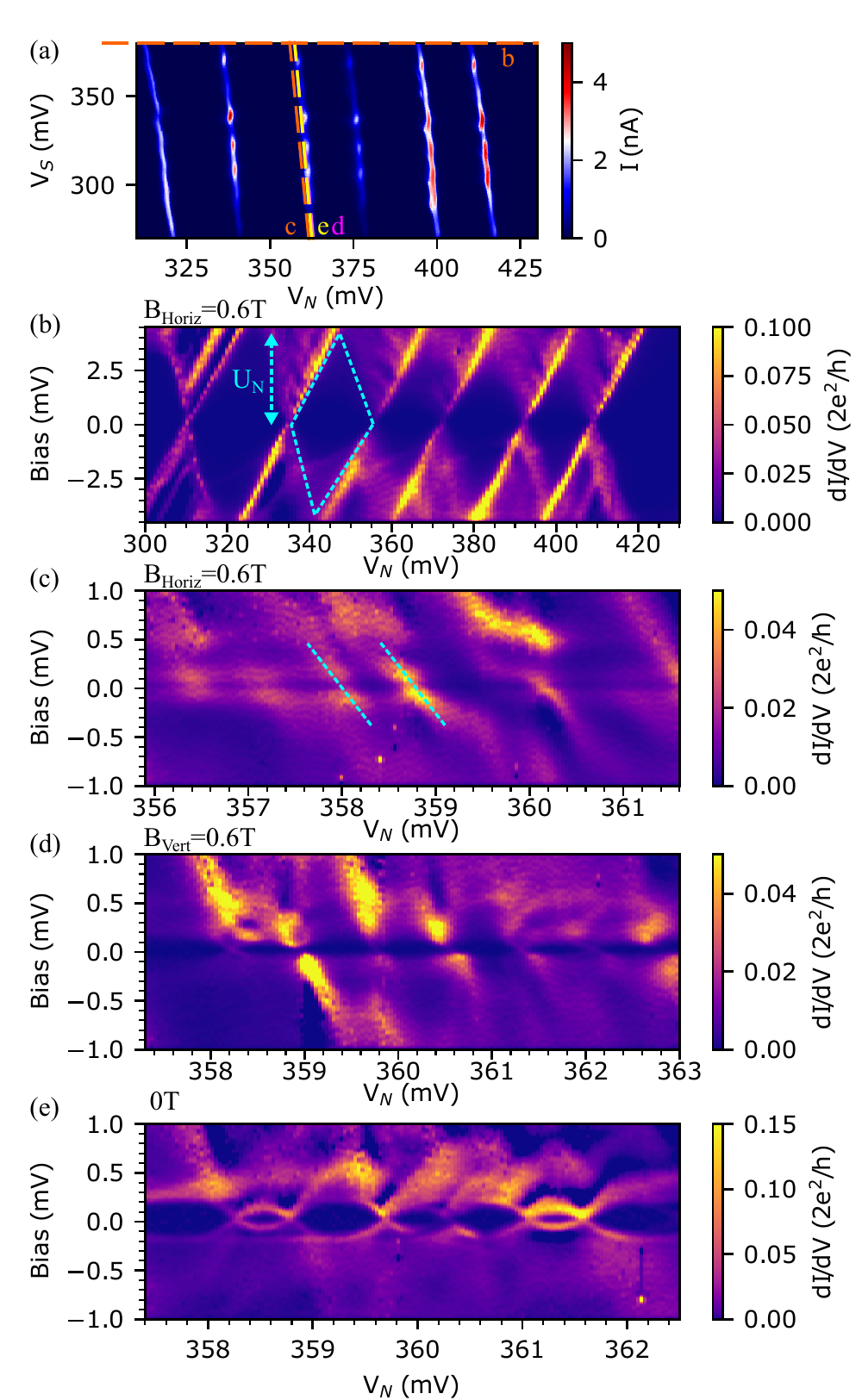}
\caption{\label{S4}
(a) This stability diagram is the same as Fig. \ref{S3}(b). The dashed lines are traces along which bias-gate spectra in panels (b-e) are taken. The magnetic fields are noted for each bias-gate panel. $B_{Vert}$ is the magnetic field used in the main text. $B_{Horiz}$ is used only in this figure. $B_{Horiz}$ is not along the wire and is not in the plane of the chip, but it is perpendicular to $B_{Vert}$. The Coulomb diamond (blue dashed diamond) in panel (b) yields a charging energy of roughly 4 mV in $\mathrm{QD_N}$. Determining the charging energy in $QD_S$ is difficult. The blue dashed lines in panel (c) show edges of Column diamonds with negative slopes. The positive slope edges are hard to distinguish except perhaps near $V_N = 360$ mV. This is because in our device the $\mathrm{QD_S}$ has a shallow barrier to the superconducting lead causing strong asymmetry. We choose the charging energy in $QD_S$ to be 0.3-0.7 meV for numerical simulations.
}
\end{figure}

\begin{figure}[H]\centering
\includegraphics{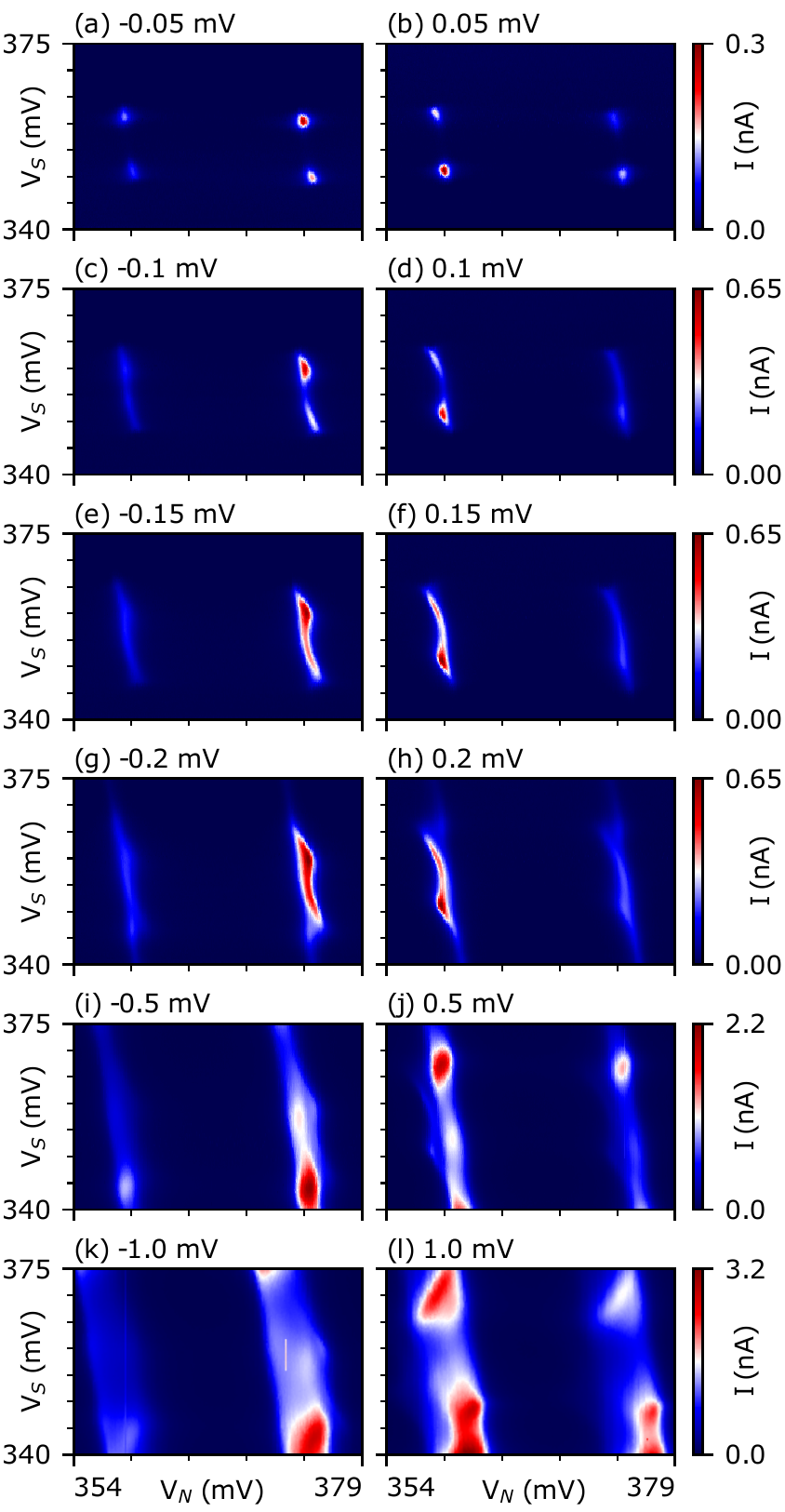}
\caption{\label{S5}
Additional data to Fig. 5, at zero magnetic field. Bias voltages are noted on each panel. At the highest bias voltages (1.0 mV) the large triangular patterns characteristic of normal (non-superconducting) quantum dots are visible in panel (l). The shapes of the triangles are different than at subgap biases, e.g. panels (d) and (f). While in this figure the bias asymmetry in the overall current appears to persist at high biases above the gap, this is not so in the larger range presented in Fig. \ref{S3}.
}
\end{figure}

\begin{figure}[H]\centering
\includegraphics{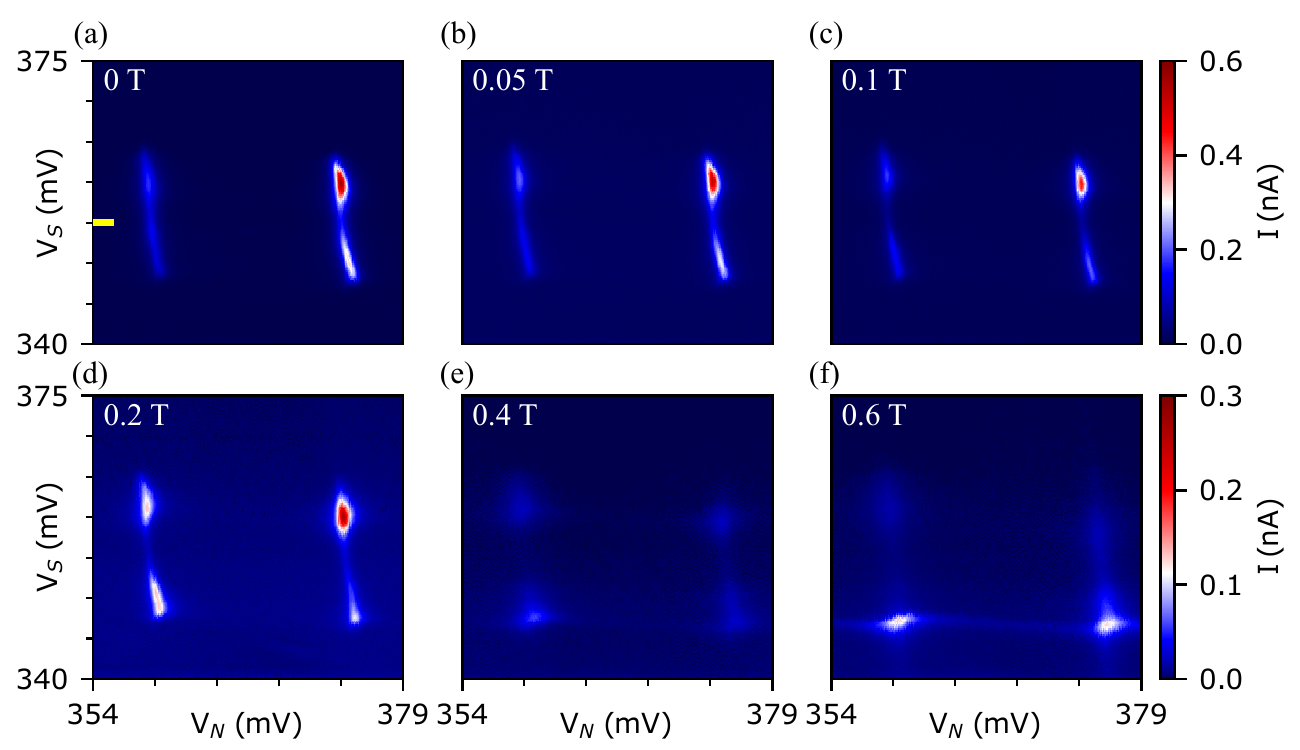}
\caption{\label{S6}
Additional data to Fig. 5, at -0.1 mV bias and different fields. The yellow short line in panel (a) indicates $V_S = 355$ mV (see Fig. \ref{S7}).
}
\end{figure}

\begin{figure}[H]\centering
\includegraphics{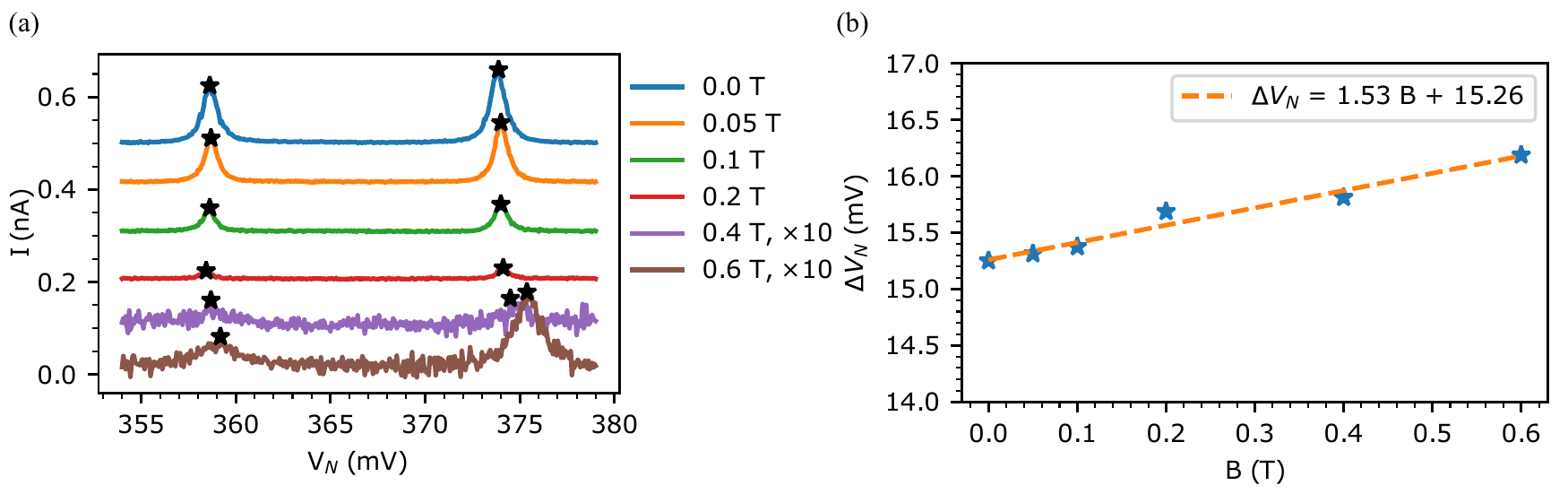}
\caption{\label{S7}
(a) Current vs $V_N$ extracted from data in Fig. \ref{S6}, at $V_S = 355$ mV. Curves are shifted vertically for clarity, with a step of 0.1 nA. Black asterisks mark resonance peaks due to levels in \ce{QD_N}. Data at 0.4 T and 0.6 T are multiplied by 10 to highlight the peaks. (b) Peak-to-peak distance vs magnetic field shows Zeeman splitting of \ce{QD_N} levels. The dashed line is a linear fit to the points. The charging energy $E_C$ of the \ce{QD_N} is about 4 mV (Fig. \ref{S4}). An effective g-factor of 6.9 can be estimated with equation $\Delta V_N = \alpha (g \mu_B B + E_C)$, where $\alpha$ is a coefficient be determined,  $\mu_B$ is the Bohr magneton, $g \mu_B B$ is the Zeeman energy. These data can be used to assign even and odd occupations in $QD_N$.
}
\end{figure}

\begin{figure}[H]\centering
\includegraphics{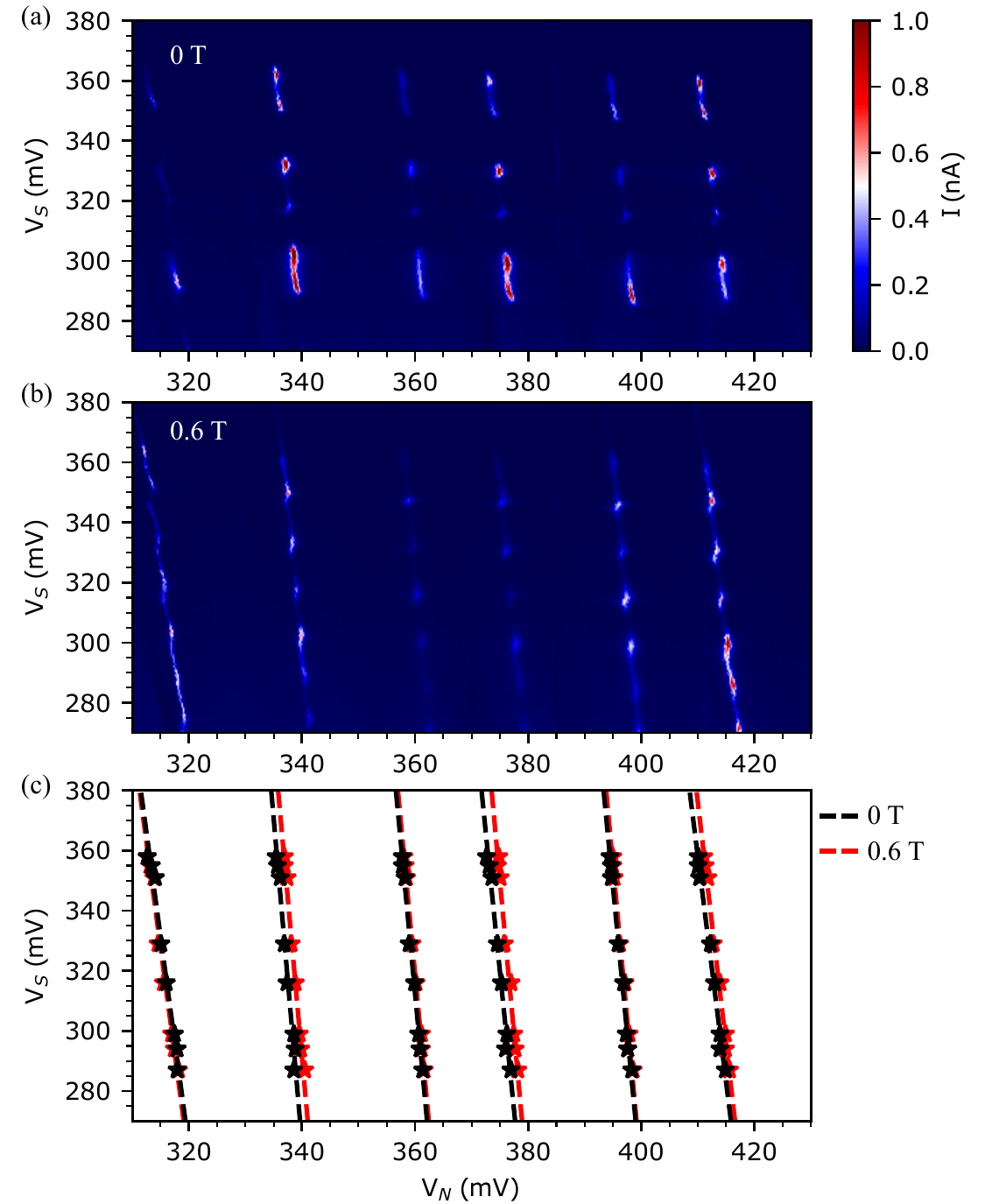}
\caption{\label{S8}
Zeeman splitting of \ce{QD_N} in the large regime discussed in the main text. (a-b) Stability diagrams at -0.1 mV. The magnetic field is noted in each panel. (c) Asterisks show peak positions extracted from a series of horizontal linecuts. Dashed lines are linear fitting lines. These data can be used to assign even and odd occupations in $QD_N$.
}
\end{figure}

\begin{figure}[H]\centering
\includegraphics{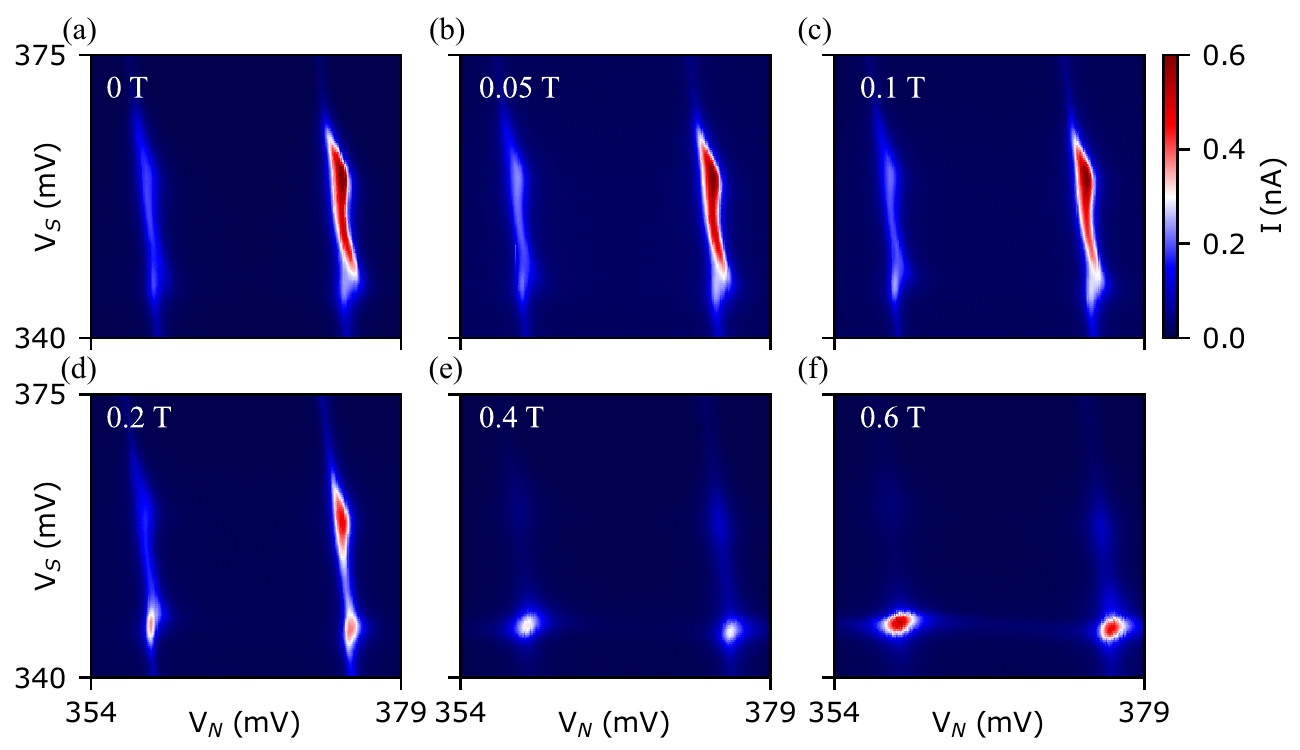}
\caption{\label{S9}
Additional data for Fig. \ref{fig5}, at -0.2 mV bias and at different magnetic fields.
}
\end{figure}

\begin{figure}[H]\centering
\includegraphics{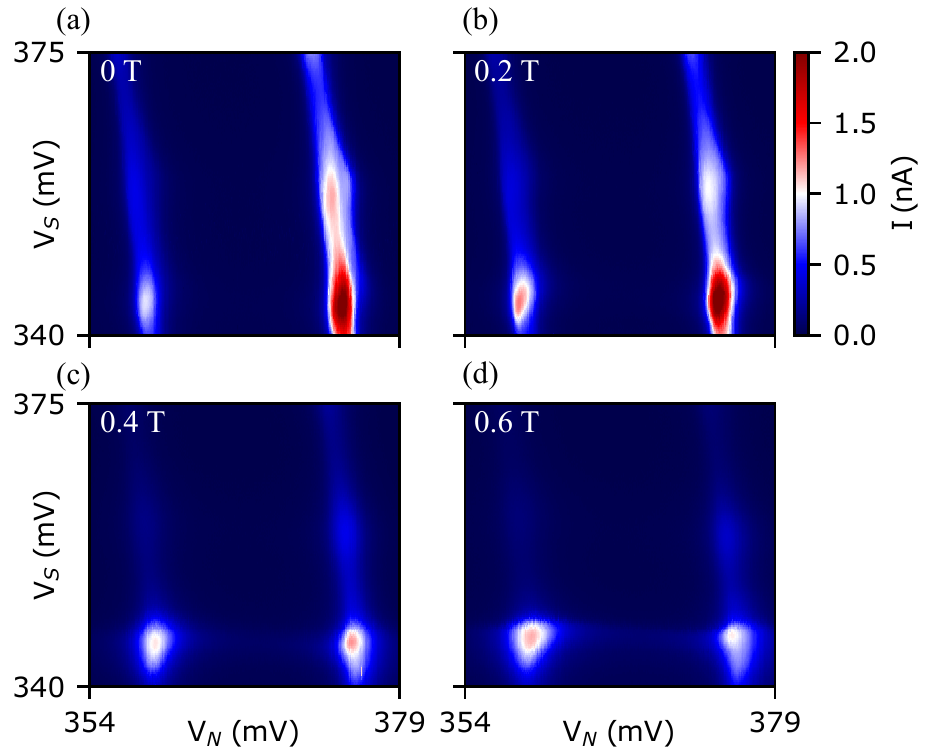}
\caption{\label{S10}
Additional data for Fig. \ref{fig5}, at -0.5 mV and different applied magnetic fields.
}
\end{figure}

\begin{figure}[H]\centering
\includegraphics{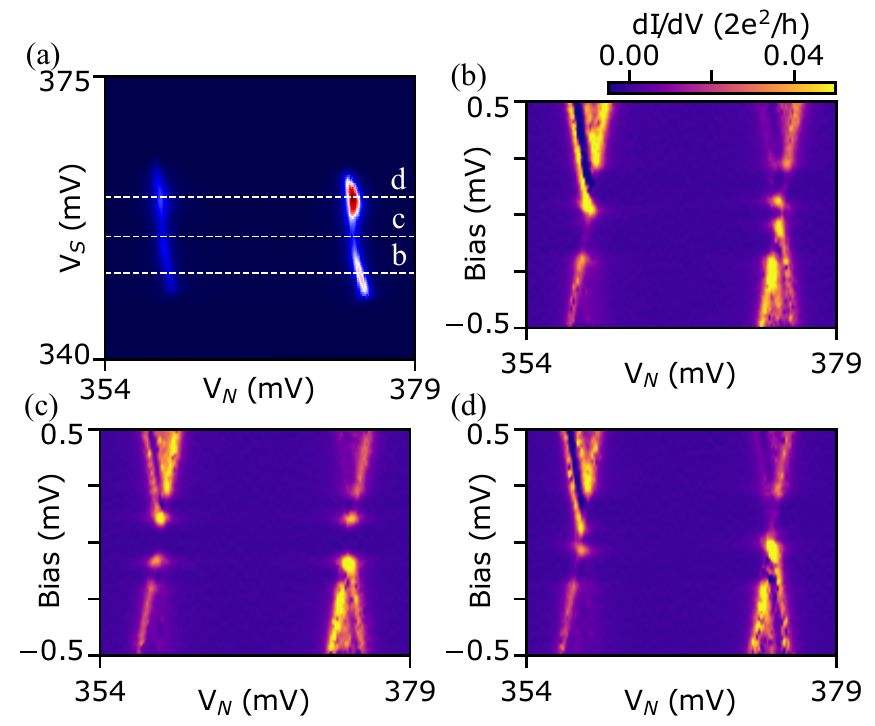}
\caption{\label{S11}
(a) This is a repeat from Fig. \ref{fig3}(a). The white dashed lines are traces along which spectra  (b-d) are taken. (b-d) Spectra at different $V_S$ values. (b,d) are the same as Fig. 2(c,d). While panel (c) is a new dataset in between (d) and (b).
}
\end{figure}

\begin{figure}[H]\centering
\includegraphics{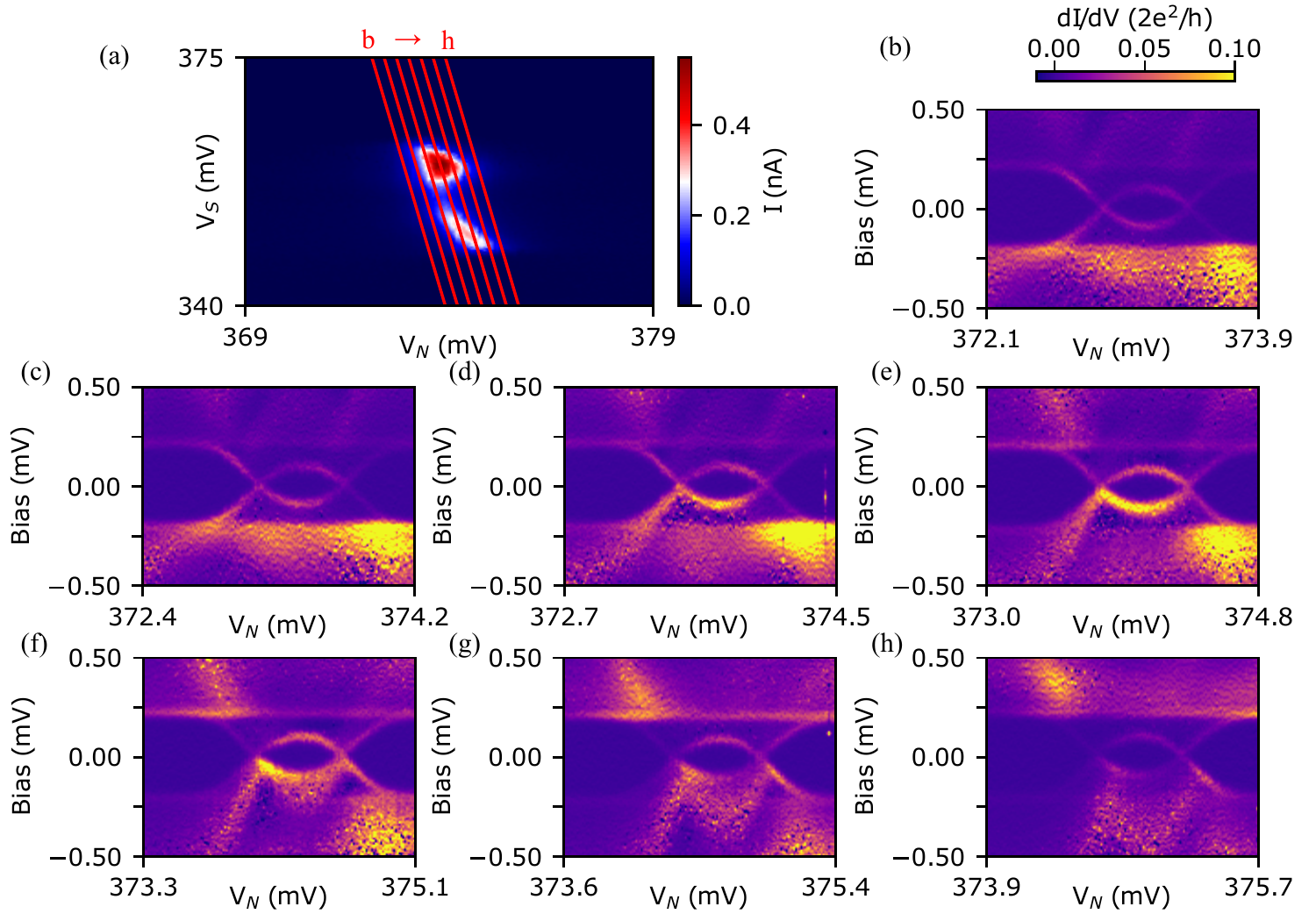}
\caption{\label{S12}
Differential conductance spectra for a series of slightly shifting traces in $V_S$-$V_N$. (a) The same data as Fig. \ref{S6}(a). Red lines are traces along which spectra (b-h) are taken. The asymmetry between positive-bias and negative-bias Andreev resonance half-loops shows up for cuts taken through the middle of the bias triangles (d-f) but is not apparent for cuts away from that regime.
}
\end{figure}

\begin{figure}[H]\centering
\includegraphics[width=\textwidth]{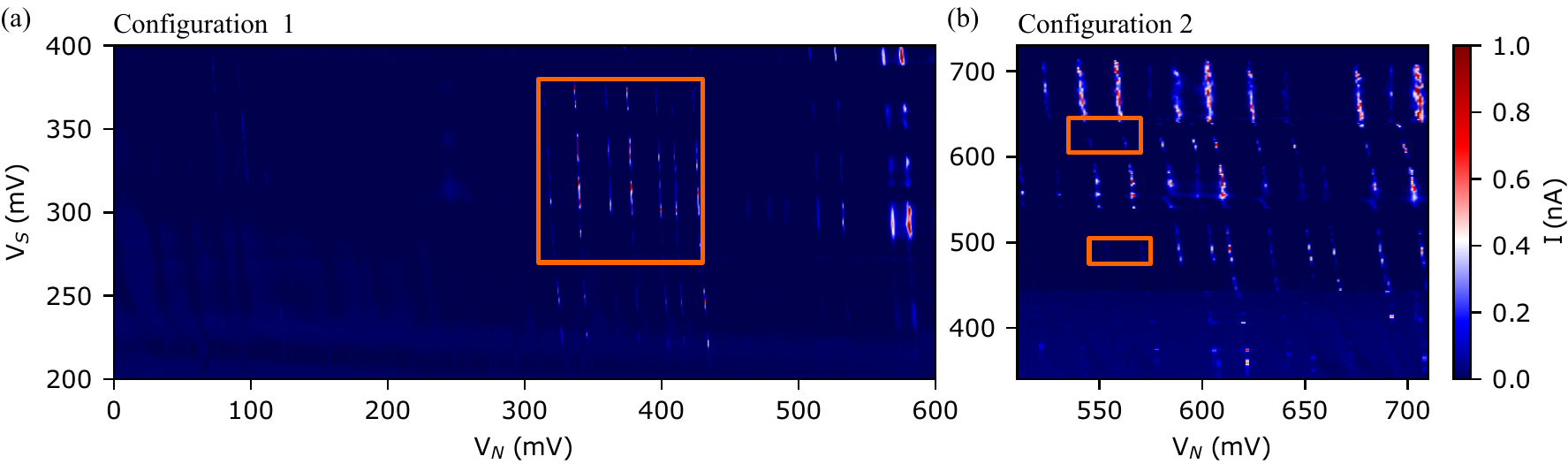}
\caption{\label{S13}
(a) The rectangle shows the regime in Fig. \ref{fig4}. There is a charge jump near \ce{V_N} = 407 mV due to the instability caused by the large range scan. (b) The same device in another gate voltage configuration obtained by re-tuning all six gates. This panel contains data from seven small-range scans. The rectangles indicate regimes where datasets in Fig. \ref{S14} and Fig. \ref{S15} are taken.
}
\end{figure}

\begin{figure}[H]\centering
\includegraphics{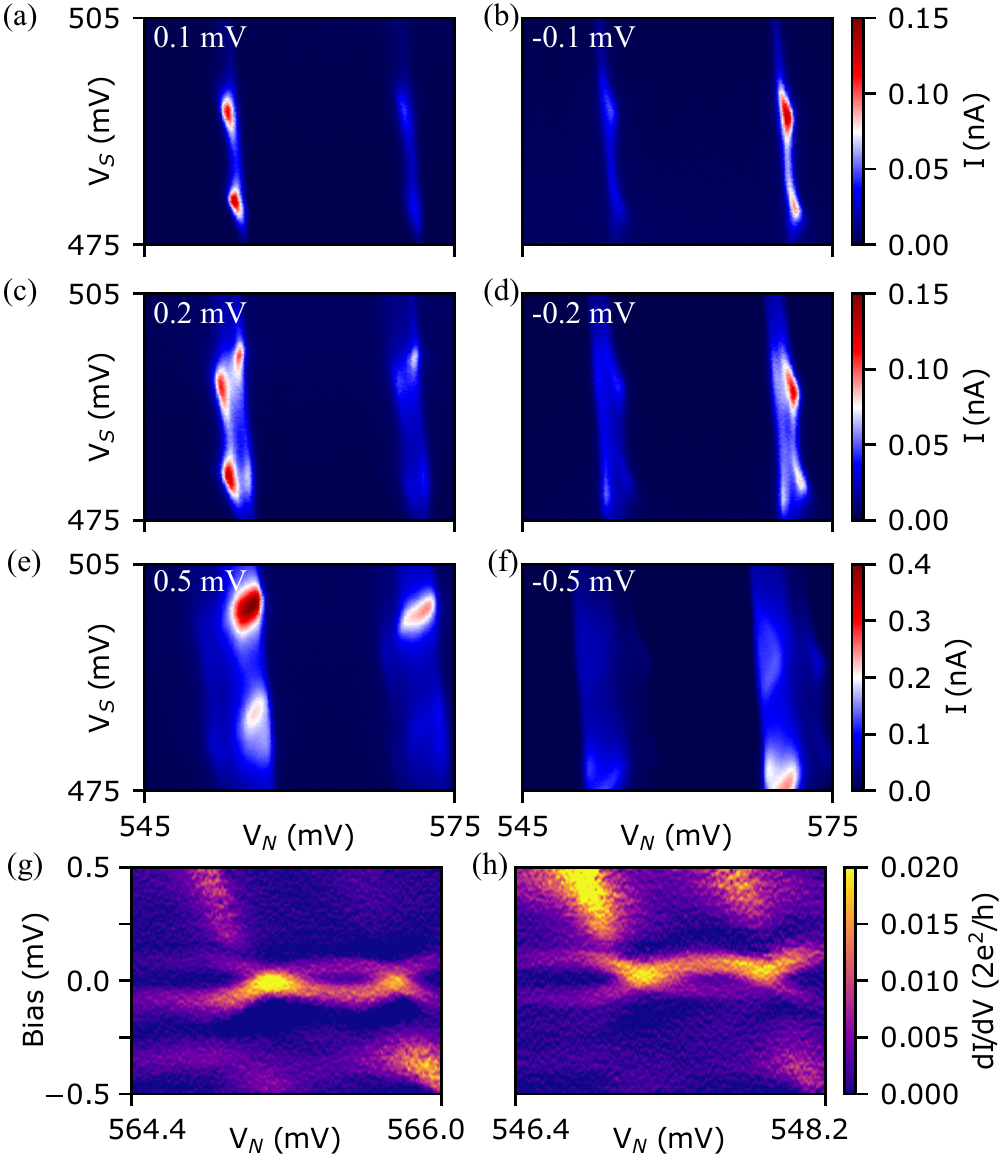}
\caption{\label{S14}
A different regime with a different gate-voltage configuration (see Fig. \ref{S13}). The stability diagrams show bias asymmetry that is similarly consistent with Andreev blockade. However in this regime the pattern does not clearly repeat over multiple periods as it does in Fig. \ref{fig4}. (a-f) The bias is noted in each panel. Spectra of \ce{QD_S} are taken when \ce{QD_N} is at (g) right and (h) left degeneracy point.
}
\end{figure}

\begin{figure}[H]\centering
\includegraphics{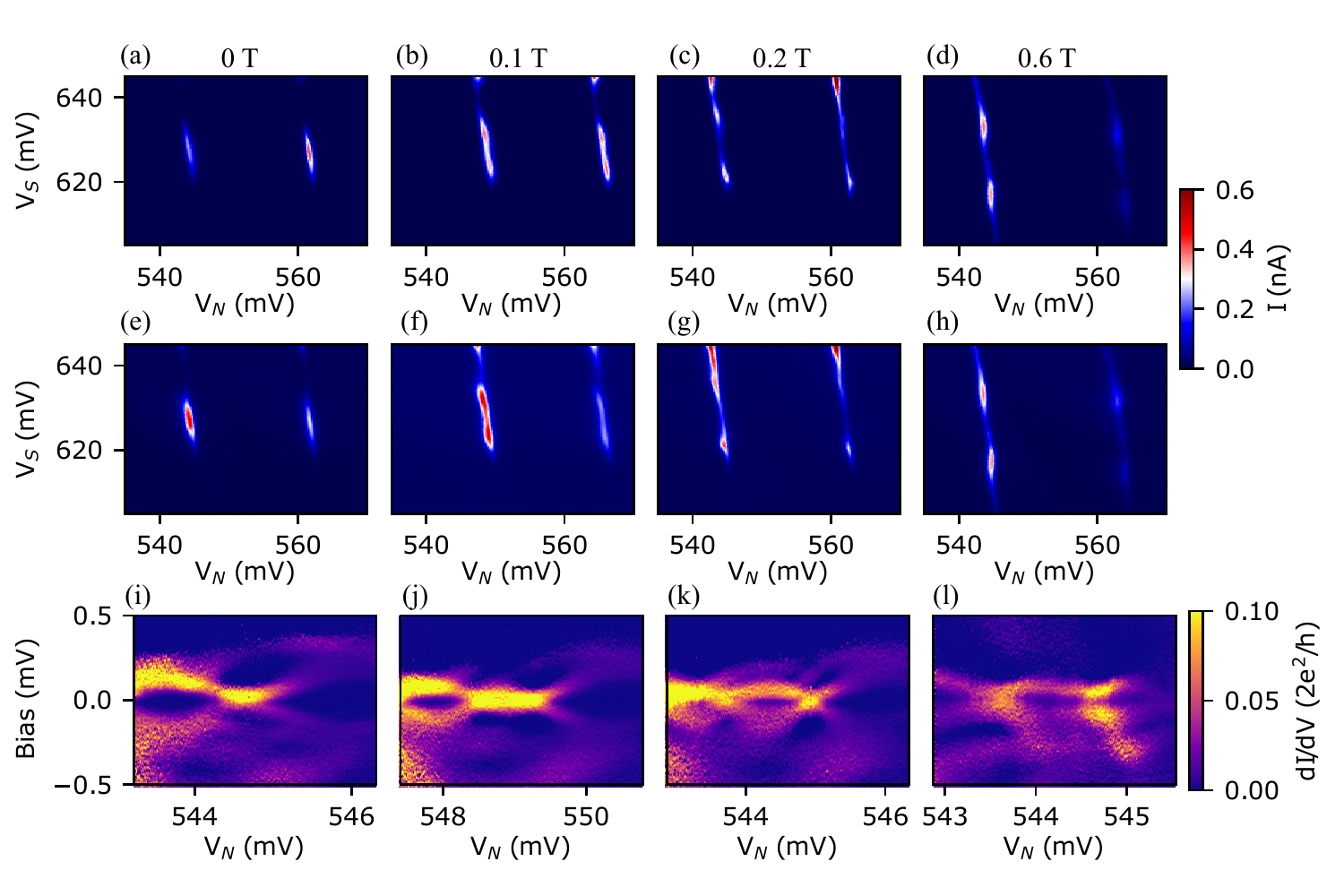}
\caption{\label{S15}
A different regime with a different gate-voltage configuration (see Fig. \ref{S13}). The stability diagrams show bias asymmetry consistent with Andreev blockade at B=0.  The asymmetry does not obviously repeat in adjacent degeneracy points. The dot $QD_S$ is stronger coupled to the superconducting lead such that the Andreev loop containing the odd-parity region is reduced to a point and we observe a single degeneracy point on the left and a single on on the right at zero magnetic field. At finite field, the pattern of four degeneracy points is restored as $QD_S$ undergoes a quantum phase transition and the odd-parity region develops. (a-d) -0.05 mV. (e-h) 0.05 mV. Spectra of \ce{QD_S} are taken along traces near \ce{QD_N}'s 0-1 transition.
}
\end{figure}

\section{Data from other devices (B and C)}

\begin{figure}[H]\centering
\includegraphics[width=\linewidth]{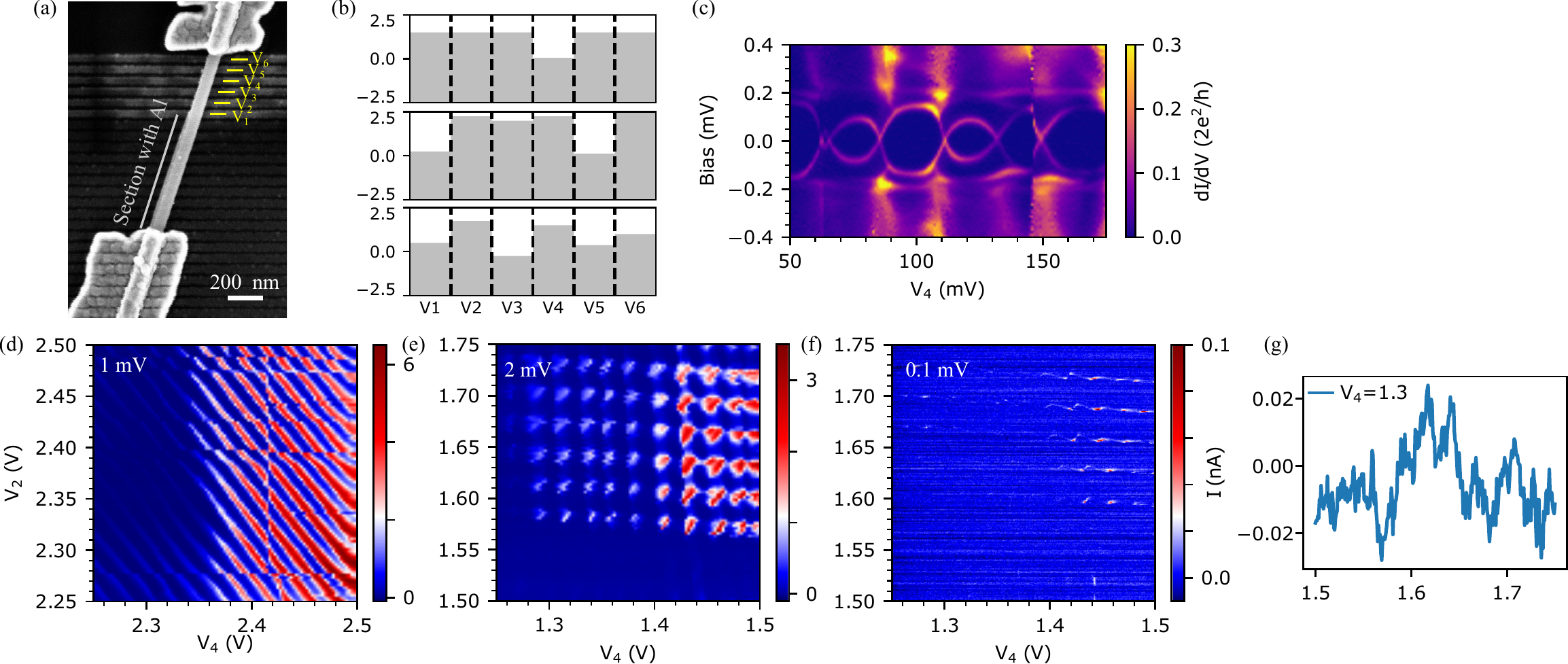}
\caption{\label{S16}
Characterization of Device B. (a) The SEM image. (b) Gate voltage configurations for datasets in panel c (top), d (middle), and e-f (bottom). The device can be tuned from a single dot regime (panel c, spectrum, and panel d, stability diagram) to a double dot regime (panel e-f). The bias voltage is indicated in white in each stability diagram. A background variation in (f) is subtracted by the current at $V_4 = 1.3$ V, which is shown in (g), for clarity.
}
\end{figure}

\begin{figure}[H]\centering
\includegraphics{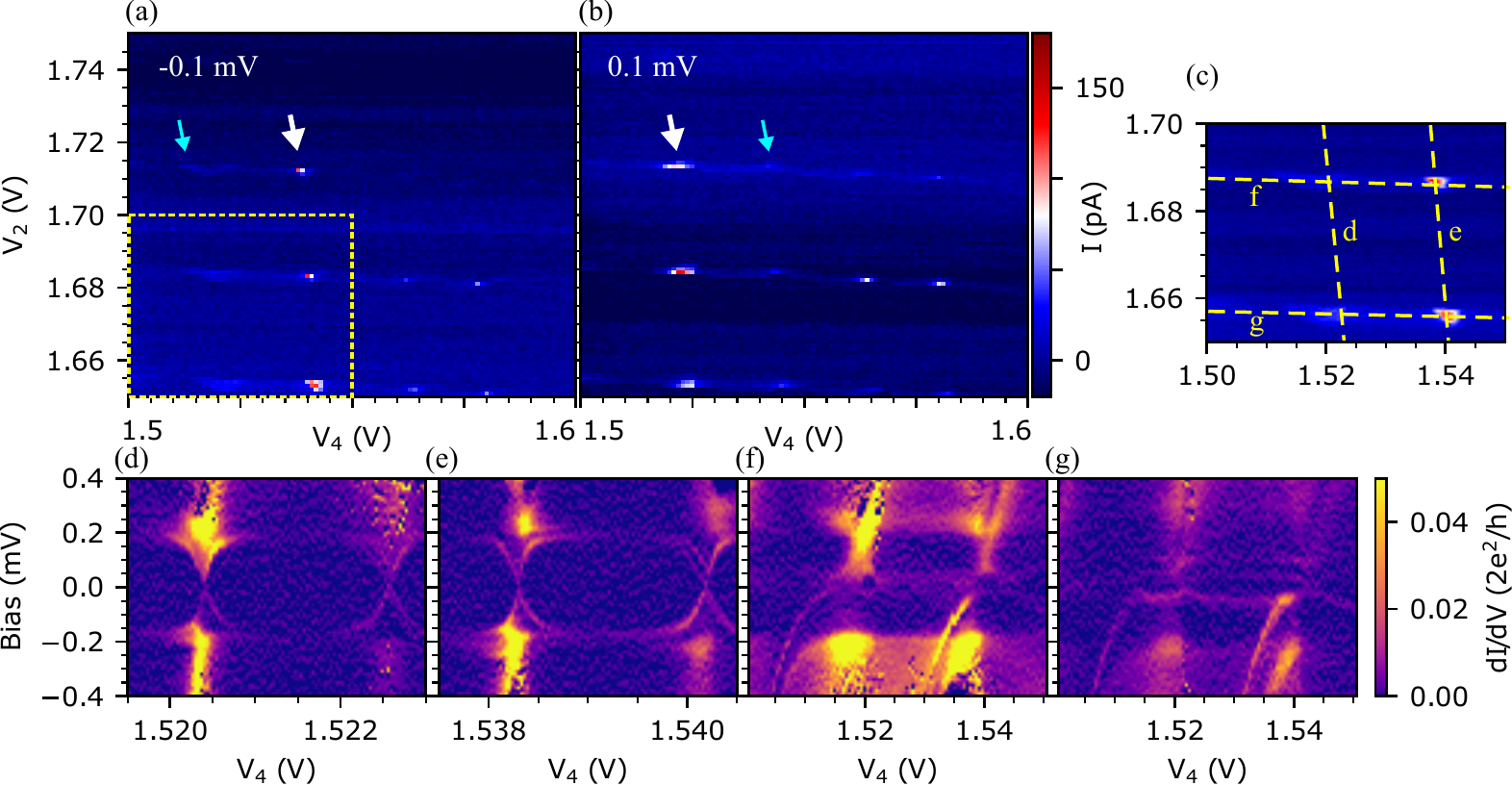}
\caption{\label{S17}
(a-b) Device B in a regime that shows bias asymmetry. The large white and small blue arrows indicate columns that have high and low currents. The pattern does not obviously repeat itself along the $V_4$ direction. We can not conclude whether the bias asymmetry is due to a coincidence or due to Andreev blockade.  (c) The rectangle-enclosed regime in panel a. Dashed lines are traces along which spectrum (d-g) are taken. Andreev bound states in \ce{QD_S} have weak and squared-shape loops (d-e), due to relatively large charging energy.
}
\end{figure}

\begin{figure}[H]\centering
\includegraphics{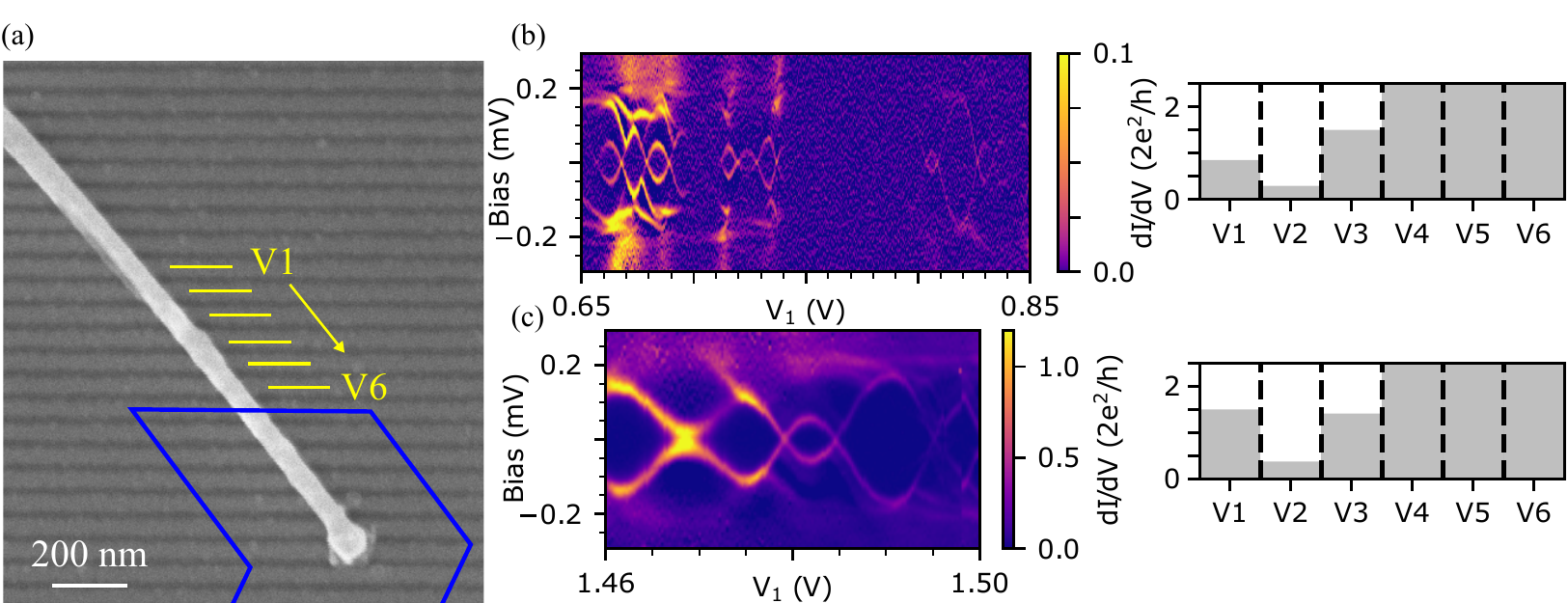}
\caption{\label{S18}
Characterization of Device C. (a) The SEM picture after etching. The blue polygon is the design pattern for the normal lead. (b-c) Spectra taken by tuning $V_1$ Andreev bound states. The gate voltage configurations are shown on the rightmost column.
}
\end{figure}

\begin{figure}[H]\centering
\includegraphics{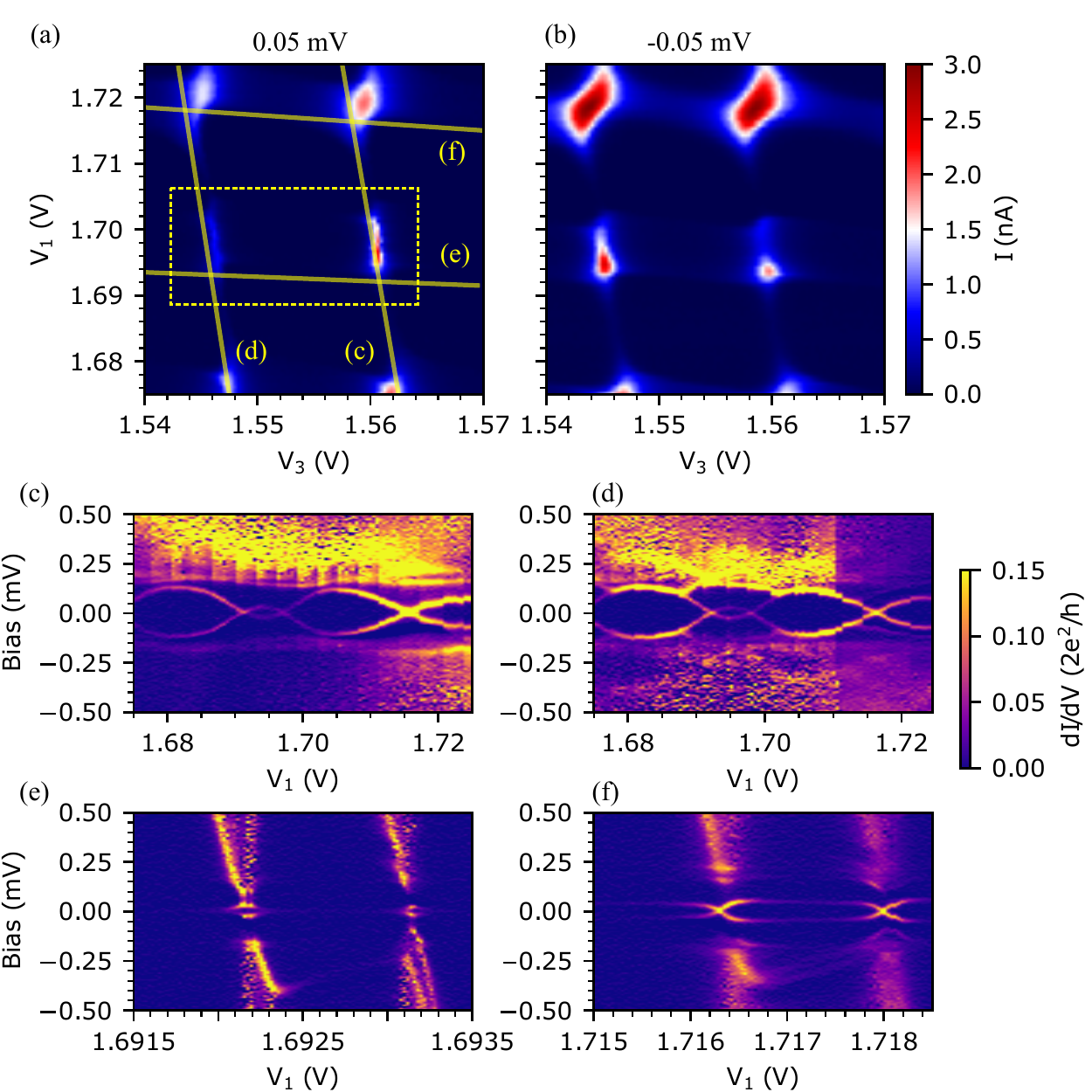}
\caption{\label{S19}
(a-b) Stability diagrams of device C. Yellow lines are traces along which the spectra in panels (c-f) are taken. The dashed rectangle highlights a regime where there is bias asymmetry.
}
\end{figure}

\end{document}